\documentstyle[12pt,ccgrra]{article}

\newcommand{\ba}{\begin{array}}
\newcommand{\ea}{\end{array}}

\newcommand{\beq}{\begin{equation}}
\newcommand{\eeq}{\end{equation}}
\newcommand{\bea}{\begin{eqnarray}}
\newcommand{\eea}{\end{eqnarray}}
\newcommand{\beal}{\setcounter{letter}{1} \begin{eqnarray}}
\newcommand{\eeal}{\addtocounter{equation}{1} \end{eqnarray}}

\newcommand{\req}[1]{Eq.(\ref{#1})}

\newcommand{\larrow}{\,\,\,\,\hbox to 30pt{\rightarrowfill}
\,\,\,\,}
\newcommand{\slarrow}{\,\,\,\hbox to 20pt{\rightarrowfill}
\,\,\,}

\newcommand{\half}{{1\over2}}

\begin{document}


\title{SINE-GORDON SOLITONS AND BLACK HOLES}

\author{J. Gegenberg\dag\footnote{e-mail:lenin@math.unb.ca},
G. Kunstatter\ddag\footnote{e-mail:gabor@theory.uwinnipeg.ca}}.
\affil{\dag Dept. of Mathematics and Statistics,
University of New Brunswick\\
Fredericton, New Brunswick, CANADA E3B 5A3}
\affil{\ddag Dept. of Physics and Winnipeg Institute of
Theoretical Physics, University of Winnipeg\\
Winnipeg, Manitoba, CANADA R3B 2E9}

\beginabstract
In the talk, the relationship between black holes in Jackiw-Teitelboim(JT)
dilaton gravity and solitons in sine-Gordon field theory is described.
The
well-known connection between solutions of the sine-Gordon equation and
constant curvature metrics is reviewed and expanded. In particular, it is shown that solutions to
the dilaton field equations for a given metric in JT theory also solve
the sine-Gordon equation linearized about the corresponding soliton.
Next, it is shown that from the B${\ddot a}$cklund transformations
relating different soliton solutions, it is possible to construct a flat
SL(2,R) connection which forms the basis for the gauge theory formulation of
JT dilaton gravity.  Finally, the supersymmetric generalization of sine-Gordon 
theory is reviewed, leading to a new approach to the problem of the origin 
of low dimensional black hole entropy.
\endabstract

\section{INTRODUCTION}

Black holes emerge from classical gravity theory, but present paradoxes
which challenge our understanding of fundamental (quantum) physics.  It
seems that the resolution of this will likely lead to a new and deeper
understanding of such issues as the unification of interactions and the
origin of elementary particles and their symmetries.  Indeed, the recent
progress in understanding black hole thermodynamics from the behaviour of
string and membrane phyics has opened new doors (`M-theory') for understanding
difficult problems in the relation of different string theories to each other
\cite{hor}.

In this paper, we begin a discussion which we hope will clarify some of
the issues raised in the attempt to understand higher dimensional black
holes in terms of non-perturbative string theory.  We will examine the
simplest {\it low dimensional} black holes in terms of an underlying
non-linear, but integrable, field theory.  In particular, we will show
that constant negative curvature two (or three) dimensional black holes
can be realized as solitons of the sine-Gordon equation and
 provide the details of this realization.  In subsequent papers we
will discuss the consequences for black hole thermodynamics and for such
problems as black hole collisions.

It has been known for a long time \cite{sol} that the solutions of the
sine-Gordon
equation
\beq
-\partial_t^2 u+\partial_x^2u =m^2\sin{ u},\label{sg}
\eeq
determine Riemannian geometries with constant negative curvature $-2m^2$ whose
metric
is given by the line-element
\beq
ds^2=\sin^2\left({ u\over2}\right)dt^2+\cos^2\left({u\over2}\right)dx^2.
\eeq
The angle $u$ describes the embedding of the manifold into a three
dimensional Euclidean space. \cite{sol}

It is also easy to show \cite{cramp} that {\it
Lorentzian} geometries with line-element
\beq
ds^2=-\sin^2\left({u\over2}\right)dt^2+\cos^2\left({u\over2}\right)dx^2,
\label{param}\eeq
have constant negative curvature if and only if
\beq
\Delta u:=\partial_t^2u+\partial_x^2u =m^2\sin{u}.
\label{esg}
\eeq

The sine-Gordon equation is a well-studied system in mathematical physics
\cite{sol}.  In particular, it is known that it is an integrable equation,
admitting solitonic solutions.  It has a rich dynamical structure, both
classically and quantum mechanically.

Recently, there has been a great deal of interest in constant curvature
black holes in two and three spacetime dimensions. The BTZ black hole
is obtained from  anti-deSitter space in 2+1 dimensions by doing suitable
identifications \cite{btz} that give the spacetime
 the global structure of a black hole with a single bifurcative horizon.
By imposing axial symmetry on 2+1 gravity, one obtains Jackiw-Teitelboim(JT)
dilaton gravity\cite{jt}: a theory of gravity
in 1+1 dimensions in which a scalar field (the dilaton) is non-minimally
coupled to the spacetime metric. The dilaton is essentially the circumference
 of the circle through a given point in the direction of axial symmetry.
 The black hole solutions
in JT gravity are therefore dimensionally reduced BTZ black holes.
In the following we will explore the relationship between the dynamics
of JT dilaton/gravity and sine-Gordon field theory.

The relationship between constant
curvature metrics and sine-Gordon solitons has been known for some time.
Moreover, our analysis is very similar in spirit to the original
one by Jackiw in Ref.\cite{jt}. In that work, the connection between Liouville
theory and constant curvature metrics in conformal gauge was used to motivate
the study of what is now called Jackiw-Teitelboim gravity.
However, besides the obvious difference of dealing with sine-Gordon theory
as opposed to Liouville theory, the present paper  goes beyond previous work
in two important respects, motivated by the current context in which
JT gravity is considered. In particular, given the importance of black hole
solutions and their relationship to the BTZ black hole\cite{btz}, the dilaton
field takes on new physical importance, beyond being simply a Lagrange
multiplier.
We will show that the dilaton field has a natural counterpart in the
sine-Gordon
theory: it is a zero-mode of the corresponding linearized
sine-Gordon equation. Secondly, we will reveal a deep connection between
B${\ddot a}$cklund transformations and the gauge theory formlulation of
JT gravity. It therefore appears that there is a duality between JT black
holes and sine-Gordon solitons that might be used to shed light on the
field theory origin of black holes and the dynamical source of black hole
entropy.

The paper is organized as follows. In Section 2 we review
Jackiw-Teitelboim gravity, with emphasis on the nature of the black hole
solutions and the  form of the gauge theory formulation of the theory\cite{bf}.
In Section 3 we outline the properties of Euclidean sine-Gordon solitons, while
 the relationship between the one-soliton solution and JT black holes is
derived in Section 4.  Section 5 shows how the
gauge theory form of JT gravity arises naturally from a consideration of the
B${\ddot a}$cklund transformations of sine-Gordon solitons. In Section 6, 
we discuss the supersymmetric extension of sine-Gordon theory, while 
Section 7 briefly explores the appearance of the dilaton field in 
{\it quantum} sine-Gordon theory.  Finally, Section 8
 closes
with conclusions and prospects for future work.

\section{JACKIW-TEITELBOIM DILATON/GRAVITY}

The simplest theory of two dimensional gravitation was first discussed in
1984 by Jackiw and Teitelboim \cite{jt}.  The equation of motion was
$R(g)=-2m^2$, i.e. that the spacetime $M_2$ had a Lorentzian
metric $g_{\mu\nu}$
with constant negative (Ricci scalar $R(g)$) curvature $-2m^2$.  In order to
derive this from
an action principle, one needs to introduce an additional spacetime
scalar field $\tau$,  with action functional
\beq
I_{JT}[\tau,g]={1\over2G}\int_{M_2}d^2x\sqrt{-g}\tau\left(R+2m^2\right),
\label{JT}
\eeq
where $G$ is the gravitational coupling  constant, which in two dimensional
spacetime is dimensionless.  Sufficient conditions that this
functional be stationary under arbitrary variations of the dilaton and
metric fields are, respectively
\bea
R+2m^2=0;\\
\left(\nabla_\mu\nabla_\nu-m^2g_{\mu\nu}\right)\tau=0.\label{jteqms}
\eea

Since the constant curvature metrics are maximally symmetric, there
are three Killing vector fields.  This follows rather directly from
the dilaton equations of motion \cite{obs} in that the three
functionally independent solutions $\tau_{(i)}, i=0,1,2$ of
\req{jteqms} determine three functionally independent vector fields $k^\mu
_{(i)}$ via
\beq
k^\mu_{(i)}={\epsilon^{\mu\nu}\over m\sqrt{-g}}\partial_\nu\tau_{(i)},
\label{kv}
\eeq
where $\epsilon^{\mu\nu}$ is the permutation symbol.  These three
vector fields satisfy the Killing equations by virtue of \req{jteqms}.
Associated with each solution and corresponding
 Killing vector there is a conserved
charge for the system, namely \cite{obs}:
\beq
M_{(i)}=-{1\over m^2}\mid\nabla\tau_{(i)}\mid^2+\tau_{(i)}^2
,\label{adm}
\eeq
When the Killing vector is timelike, it can be shown that this corresponds
to the ADM energy of the solution.

Although all the solutions of Jackiw-Teitelboim gravity are {\it locally}
diffeomorphic to two-dimensional anti-DeSitter spacetime, one may obtain
distinct {\it global} solutions, some of which display many of the
attributes of black holes \cite{obs,robb1}.  For example, consider a solution
where the metric is given by the
line-element
\beq
ds^2=-\left(m^2 r^2-M\right)dt^2+\left(m^2 r^2-M\right)^{-1}dr^2,\label{bh}
\label{eq: black hole metric}
\eeq
where $M$ is a constant and the dilaton is
\beq
\tau=mr.
\label{eq: dilaton solution 1}
\eeq
This metric is a dimensionally truncated three dimensional BTZ black hole
\cite{btz}.  Clearly there is an event horizon located
at $r=\sqrt{M}/m$.  It is important to note here that though this fact
can be easily read off from the metric, since the latter is in manifestly
static form, it also follows from solving for the variable $r$ in the
equation $\mid k^\mu\mid^2:=g_{\mu\nu}k^\mu k^\nu=0$, where $k^\mu$ is
the Killing vector field determined by the dilaton $\tau$ above via
\req{kv}. It is straightforward to show that the ADM energy of the
black hole solution \req{eq: black hole metric} and
\req{eq: dilaton solution 1}
 is
\beq
E_{BH}=mM/2G,
\eeq

  We note for later reference that there is a conical singularity located in
the BTZ black hole at $r=0$. In the context of JT gravity there
is no conical singularity, but the vanishing of the
dilaton field gives rise to an infinite effective Newton's constant. Surfaces
for which $\tau=0$ should therefore be excluded from the manifold.

It was mentioned above that for any given metric, there exist three linearly
independent solutions to the dilaton field equations. For
the metric given in \req{eq: black hole metric} the simplest solution
is the standard one (\req{eq: dilaton solution 1}). The other two solutions
for this metric are:
\bea
\tau_{(2)} &=& \sqrt{m^2r^2 - M}\sinh{\sigma}+ constant
\label{eq: dilaton solution 2}\\
\tau_{(3)} &=&  \sqrt{m^2r^2 - M}\cosh{\sigma} + constant
\label{eq: dilaton solution 3}
\eea
where  $\sigma\equiv m\sqrt{M}t$. The corresponding
Killing vectors are:
\bea
{\vec k}_{(2)} &=& \left({mr\over\sqrt{m^2r^2-M}}\sinh\sigma,
  -\sqrt{M}\sqrt{m^2r^2-M} \cosh\sigma\right)\\
{\vec k}_{(3)} &=& \left({mr\over\sqrt{m^2r^2-M}}\cosh\sigma,
  -\sqrt{M}\sqrt{m^2r^2-M} \sinh\sigma\right)
\eea
A straightforward calculation shows that the conserved charge \req{adm} is in
fact
the same for all values of the dilaton:
\beq
M_{(1)}=M_{(2)}=M_{(3)}= M
\eeq
providing that the constants in \req{eq: dilaton solution 2} and \req{eq:
dilaton solution 3} are set to zero.
\par
There is also a gauge theory version of Jackiw-Teitelboim gravity \cite{bf}.
The action functional is
\beq
I_{BF}[\phi,A]={1\over2G}\int_{M_2}\,\,Tr\left(\phi F(A)\right),\label{BF}
\eeq
where $\phi$ is a spacetime scalar with values in the Lie algebra {\it
sl}(2,$R$), $A$ is an SL(2,$R$) connection on a principal bundle over
spacetime $M_2$, whose  curvature is denoted by $F(A):=dA+\half[A,A]$.
The trace is over the adjoint representation of SL(2,$R$).

The stationary configurations of $I_{BF}$ are
\bea
F(A)=0,\\
D_A\phi:=d\phi+[A,\phi]=0.
\eea

It is well-known \cite{bf} that if  we identify two of the components, say
${1\over m}A^a, a=0,1$  with the spacetime frame-field $e^a$, and the
remaining component, $A^2$ with the spin-connection $\omega$, then the
zero-curvature  condition for $A$ implies that the connection $\omega$
is torsion-free,
\beq
D_\omega e^a:=de^a-\epsilon^a{}_b\omega\wedge e^b=0,
\eeq
and has constant negative curvature
\beq
d\omega-m^2\epsilon_{ab}e^a\wedge e^b=0.
\eeq
Note that our convention here is that $A:=A^iT_i$ where the $T_i$ are the
generators of SL(2,$R$) obeying
\beq
[T_i,T_j]=\epsilon_{ijk}\eta^{kl}T_l,
\eeq
where $\eta^{ij}$ is the Minkowski metric with signature $+2$ and $\eta^{00}=
-1$ and $\epsilon_{ijk}$ is the permutation symbol.

If we define the metric tensor
\beq
g_{\mu\nu}:=\eta_{ab}e^a_\mu e^b_\nu,
\eeq
then the Ricci scalar curvature satisfies $R(g)=-2m^2$.  Hence every solution
of the geometrodynamic version is a solution of the gauge theory version,
but it is easy to see that there are solutions of the gauge theory, (e.g.
$A=0$) which do not immediately yield {\it non-degenerate}
Lorentzian geometries of constant
negative curvature.

We note here that the information about the dilaton and Killing vector
fields is encoded in the covariant constancy of the Lie-algebra valued
scalar field $\phi=\phi^iT_i$.  Up to a constant multiple, one may
identify the dilaton with $\phi^2$, and the frame-field components
of the corresponding Killing vector field with the $\phi^a$.

\section{THE SINE-GORDON EQUATION AND CONSTANT CURVATURE GEOMETRY}

We have seen that when the Lorentzian metric is parametrized as in
\req{param}, it has constant negative curvature $-2m^2$ if the function
$u(t,x)$ satisfies the Euclidean sine-Gordon equation \req{esg}.  In the
following we summarize the properties of this non-linear partial differential
equation, specializing to the slightly less well-studied case of Euclidean
signature \cite{sol}.

The only obvious solution of the sine-Gordon equation is the trivial
solution $u_n(t,x)=2\pi n$.  These solutions are important because, though
locally trivial, they can be used, in conjunction with B${\ddot a}$cklund
transformations, to generate an infinite family of non-trivial soliton
solutions.  The B${\ddot a}$cklund transformations are a one parameter family
of first order non-linear partial differential equations in two real-valued
functions $u,u'$, whose integrability conditions are precisely the
sine-Gordon equations for $u$ and $u'$.  We write the B${\ddot a}$cklund
transformations in the form \cite{akns,sol}:
\bea
\Gamma_{,z}&=&\half\left(1+\Gamma^2\right)u_{,z}+\half m k \Gamma,\\
\Gamma_{,\bar z}&=&{m\over2k}\left[\cos{u} \Gamma-\half\sin{u}\left(1-\Gamma^2
\right)\right].\label{back}
\eea
In the above, $\Gamma$ is defined by
\beq
\Gamma:=\tan\left({u+u'\over4}\right);
\eeq
the quantity $z:=x+it$, with $\bar z$ its complex conjugate. Finally,
$k$ is a  complex parameter, in general.  One can verify that the
integrability condition $u_{,xt}=u_{,tx}$ implies that $u'$ satisfy
\req{esg}, and similarly exchanging $u$ with $u'$.

Starting with the trivial seed solution $u=0$, it follows from \req{back}
that
\beq
u'=4\tan^{-1}\exp\left\{\pm m\gamma[x-x_0-v(t-t_0)]\right\},\label{1sol}
\eeq
with $\gamma:=(1+v^2)^{-\half}$,
is also a solution of \req{esg}.  In the above $t_0,x_0$ are
integration constants.  In order to obtain a real solution $u'$, the
parameter $k$ must have modulus one, i.e. $k=e^{i\alpha}$, and we
have $\cos{\alpha}=\gamma,\sin{\alpha}=v\gamma$.

The solution with the $+$ sign in the exponent is the 1-soliton
solution; the opposite sign is the anti-soliton solution.  Upon `Wick
rotation' to the Lorentzian signature, (and in this case $v\to iv$), one
sees that the soliton(anti-soliton) propogates through space with constant
velocity $v$ $(-v)$.  Hence we may think of the soliton as being located at
$x=vt$ at time $t$.

The B$\ddot a$cklund transformation may be used to generate multi-soliton
solutions.  For example, if we use the 1-soliton solution for $u$, then the
\req{back} yields the 2-soliton solution for $u'$.  This iteratively
defines an infinite family of independent solutions of the sine-Gordon
equation.  In the sense to be described below, these solutions are
topologically distinct.

The following vector field
\beq
j^\mu:={1\over2\pi}\epsilon^{\mu\nu}\partial_\nu u,
\eeq
(on $E_2$) is trivially conserved, i.e. $\partial_\mu j^\mu =0$ identically
for any smooth function $u$.  The {\it soliton number} of $u$ is defined
to be
\beq
Q[u]:=\int_S ds n_\mu j^\mu,
\eeq
where $S$ is a surface in $E_2$ such that $n^\mu$, its normal, is
non-spacelike.  It is easy to see that $Q[u]=\pm 1$ if $u$ is respectively,
the 1-soliton (anti-soliton).   An {\it n-soliton} $u$ has $Q[u]=n$.

Besides the n-solitons, there are topologically trivial ($Q[u]=0$), but
locally non-trivial,
solutions.  We will not discuss these here, except in the following
general sense.  Obviously a given solution $u$ of the sine-Gordon
equation can not be constructed as a linear superposition of some
basis of solutions.  In fact, the best one can do is the `inverse
scattering picture'.  Here one looks at a solution $u(t,x)$ as a
series of snapshots taken at various times $t$.  The data at $t=-\infty$
(or at $+\infty$) at least partially determines the structure, including
its topological structure, of a given $u$.

\section{THE 1-SOLITON AS A BLACK HOLE}

In this Section, we indicate the connection between the one-soliton
in sine-Gordon theory, and JT black holes.
 When the metric is parametrized as in
\req{param}, the action \req{JT} takes the form:
\beq
I_{JT}[\tau,u] = {1\over 2G} \int_{M_2}d^2x \tau (\nabla u - m^2 \sin{u}),
\eeq
where the $\nabla$ denotes the {\it flat space Euclidean} Laplacian. The field
equations that result from a variation of this ``reduced'' action are the
sine-Gordon equation \req{esg} for the field $u$ and the {\it linearized}
sine-Gordon equation for the dilaton:
\beq
\left(\Delta-m^2\cos{u}\right)\tau=0.\label{linsg}
\eeq
Note that \req{linsg} can also be derived directly from the equations of motion
 \req{jteqms} for the unreduced theory. It is interesting that the dilaton
fields
which generate the Killing vectors of the constant curvature metric in
JT gravity also correspond to symmetries of the sine-Gordon model. The dilaton
field infinitesmally maps solutions of the sine-Gordon equation on to
other solutions. That is, the
field $u'= u+\epsilon\tau$, where $u$ and $\tau$ obey \req{esg} and
\req{linsg}, also solves \req{esg} to first order in $\epsilon$. However, it
should be
noted that there are more solutions to the linearized sine-Gordon equation
than there are to the dilaton equations.

We shall now demonstrate that the 1-soliton solution \req{1sol} of the
sine-Gordon equation determines a metric in a coordinate patch on $M_2$
in which there is a Killing vector field which is timelike in the
asymptotic region and becomes null at an interior point of the patch.  In
other words, it determines a black hole metric.  Indeed, when \req{1sol} is
used in the Lorentzian metric \req{param}, the latter simplifies to:
\beq
ds^2_{1-sol}=- \hbox{sech}^2{\rho} dt^2+\tanh^2{\rho} dx^2,\label{solbh}
\eeq
where
\beq
\rho:=m\gamma(x-vt),
\eeq
and we have chosen for simplicity $x_0=t_0=0$.  We perform successive
coordinate transformations:  first from $(t,x)\to (T,\rho)$, with $\rho$
as defined above and with
\beq
dT=dt-v{\tanh^2{\rho}\over m\gamma(\hbox{sech}^2{\rho}-v^2\tanh^2{\rho})}d\rho.
\label{1solmetric}
\eeq
Next we transform from $(T,\rho)\to (T,r)$ with
\beq
r:={1\over m\gamma}\hbox{sech}{\rho}.
\eeq
Both of these transformations have non-zero Jacobian for all $(t,x)$ in
$R^2$.  The result is the metric with line-element:
\beq
ds^2_{bh}=-\left(m^2 r^2-v^2\right) dT^2+\left(m^2 r^2-v^2\right)^{-1}dr^2.
\eeq
This is the metric of a Jackiw-Teitelboim black hole with total energy
$mv^2/2G$ and event horizon at $r=v/m$.

We see that the metric \req{solbh} is Kruskal-like in that there is
no coordinate singularity at the horizon, but is singular (in fact degenerate)
at $\rho=0$, which is the location of the soliton, and at $r=0$, i.e. where 
$\rho\to\infty$.

\section{FROM SINE-GORDON TO GEOMETRY}

So far we have seen that if the coordinates of spacetime are fixed so that
the metric is of the form \req{param}, then $u$ must satisfy the sine-Gordon
equation and the dilaton $\tau$ must satisfy the linearized sine-Gordon
equation.  So we have `derived' sine-Gordon from carefully dressed
Jackiw-Teitelboim gravity theory.  We now address the question of reversing
this and deriving gravity from the structure of sine-Gordon theory.   In
fact, the work over twenty years ago by Ablowitz et. al. \cite{akns} and
others \cite{sol} provides a partial answer to this.

Consider the covariant constancy condition
\beq
D_A w:=dw+Aw=0,\label{w}
\eeq
where $A$ is an SL(2,R) connection of the form
$A=A^iT_i$ with
\bea
A^0&=&{m\over2}\left(kdz+k^{-1}\cos{u}d\bar z\right),\nonumber\\
A^1&=&{m\over2 k}\sin{u} d\bar z,\nonumber\\
A^2&=&u_{,z} dz,\label{A}
\eea
and
\beq
w:=\left[\begin{array}{clcr}w_1 \\ w_2\end{array}\right],
\eeq
Then if we perform the Ricatti transformation
\beq
\Gamma=w_2/w_1,
\eeq
it follows from \req{w} that the $u,u'$, with $\Gamma=\tan[(u+u')/4]$ are
related by a B$\ddot a$cklund transformation \req{back}.  It is also
straightforward to check that the connection $A$ is {\it flat}, i.e., that
$F(A)=0$.

There is a natural metric on $R^2$ induced by the flat connection $A$.  This
comes from the Casimir invariant metric $\eta_{ij}:=2 Tr\left(T_i T_j\right)$,
where the generators $T_i$ of SL(2,R) are defined so that $\eta_{00}=
\eta_{11}=-\eta_{22}=1$.  To get a real Lorentzian metric on $R^2$, we must
choose the parameter $k$ in the connection to have modulus one, i.e.
$k=e^{ia/2}$.  The complex frame-field is
\beq
\{E^0,E^1\}:={1\over m}\{A^0,A^1\},
\eeq
 and the spin-connection
is
\beq
\Omega:=A^2={i\over2}\left(u_{,x}dt-u_{,t}dx\right)+\half du.
\eeq
  The line-element is {\it real}:
\bea
ds^2&:=&\left(E^0\right)^2+\left(E^1\right)^2\nonumber\\
&=&-\half(\cos{a}-\cos{u})dt^2+
\half(\cos{a}+\cos{u})dx^2-\sin{a}\,\,dtdx,\nonumber\\
\,
\eea
and has constant negative curvature given by $Im(\Omega)$ if $u$ obeys the
sine-Gordon equation.

\section{THE SUPERSYMMETRIC EXTENSION}
We summarize here the supersymmetric extension of the sine-Gordon
model and speculate about the relationship of the latter to two
dimensional gravity.  I will adhere to the conventions in the 1978
Witten-Olive paper \cite{witten}.

The supersymmetric extension of the sine-Gordon equation has the action 
functional \cite{witten}
\beq
I_{SSsG}[u,\psi]={1\over\beta^2}\int d^2x\left(\half\mid\nabla u\mid^2+
{i\over2}\bar\psi\gamma^\mu\partial_\mu\psi-\half\sin^2u+\half\cos u\,\bar
\psi\psi\right),\label{sssg}
\eeq
where $\psi(t,x)$ is a Majorana spinor field and the $\gamma^\mu$ are the 
Dirac matrices.  Clearly  when $\psi\equiv 0$, this reduces to sine-Gordon 
theory.

The supersymmetry algebra is generated by $Q_\pm$ given by
\beq
Q_\pm=\int dx\mid\left(\partial_t u\pm\partial_x u\right)\psi_\pm
\pm\sin u\psi_\mp\mid,
\eeq
where $\psi_\pm$ are the spinor components of $\psi$.  The supersymmetry 
algebra is \cite{witten}
\beq
Q_+^2=P_+;\,\,\,Q_-^2=P_-;\,\,\,\left\{Q_+,Q_-\right\}=T,\label{susyalgebra}
\eeq
where $P_\pm:=P_t\pm P_x$ with $P_\mu$ the generators of momentum.    
The operator $T$ is a central charge:
\beq
T:=\int dx\partial_x(2\cos u).
\eeq

Witten and Olive show that there is a BPS bound on the total energy $M:=P_+ +
P_-$, namely
\beq
M\geq\half\mid T\mid.
\eeq
Classically, the  bound above is saturated by the multi-soliton solutions of 
sine-Gordon equation.  The situation is not clear quantum mechanically at 
the moment.  Although Witten and Olive conjectured that in the quantum 
sine-Gordon theory the bound was saturated by soliton states, a recent 
preprint by Rebhan and van Nieuwenhuizen indicates that this is 
not true \cite{rebhan}.  It is an 
open question, then, as to which quantum states do saturate the BPS bound.

We speculate that the degeneracy in the BPS saturated states of quantum 
supersymmetric sine-Gordon theory accounts for the Bekenstein-Hawking 
entropy of the two-dimensional Jackiw-Teitelboim black holes.

\section{QUANTUM SINE-GORDON THEORY AND QUANTUM BLACK HOLES}
In this section, we show how the dilaton field $\tau$ in Jackiw-Teitelboim 
gravity emerges from the semi-classical quantization of the  
sine-Gordon theory.  

When the metric is parametrized as in 
\req{param}, then the dilaton equations of motion \req{jteqms} neccessitate 
\beq
\left(\Delta-m^2\cos{u}\right)\tau=0,
\eeq
the {\it linearized} sine-Gordon equation.  Although we do
not yet fully understand this, the following is a clue to understanding 
why the dilaton is a zero-mode of the linearized sine-Gordon operator.  
The partition function for sine-Gordon theory,
\beq
Z=\int d\mu[u]\,\,e^{-I_{SG}[u]},\label{partition}
\eeq
where $I_{SG}[u]$ is the action functional for sine-Gordon theory:
\beq
I_{SG}[u]={1\over\beta^2}\int_{E_2}d^2x\left[\half\mid\nabla u\mid^2-m^2
(\cos{u}-1)\right],\label{sgaction}
\eeq
with $\beta$ the coupling constant.  We compute the one-loop effective 
action $\Gamma_{eff}$, i.e., we expand $u$ about a classical solution $\bar u$ of the 
sine-Gordon equation, keeping only terms to second order in the 
perturbation, and integrate over the perturbation.  The result is 
\beq
\Gamma_{eff}=\ln\det{}'{\left(\Delta-m^2\cos{\bar u}\right)},\label{oneloop}
\eeq
where $\det'({\cal O})$ excludes the zero-modes of the operator ${\cal O}$.  
Thus the dilatons are the non-physical modes generated 
by the symmetries of the sine-Gordon action, which must be excluded 
from the functional determinant $\det'$.  This is consistent with the 
fact that the dilatons are `potentials' for the Killing vectors.
But there are an infinite number of solutions of linearized SG, besides 
the 3 which generate the isometries of the spacetime metric.  Work is underway 
to understand these zero-modes.

\section{CONCLUSION AND SPECULATIONS}

We have seen that the sine-Gordon theory may be used to provide an
alternative classical description of the simplest two dimensional gravity
theory.  The latter necessarily contains, besides the metric tensor,
a scalar field (the dilaton), and we have seen that such a field also
emerges naturally
from the sine-Gordon theory.  Finally, we have shown  that the black hole
solutions of JT dilaton/gravity originate from the 1-soliton solutions
of sine-Gordon theory.

At this level, what remains for us to understand is, first, the
relation between the integrability of sine-Gordon theory (which has
the concommitant of the existence of an infinite number of conserved
quantities) and the zero-modes of the linearized sine-Gordon
equation.  Second, we would like
to examine the role of multi-solitons and other non-trivial solutions of the
sine-Gordon equation in JT dilaton/gravity.

Our ultimate goal is to understand the relation  between  {\it quantum} JT
dilaton/gravity and sine-Gordon theory.
In particular, we speculate that the microstates of the
black hole responsible for the black hole entropy originate in the
degeneracy of the BPS saturated states in the quantum supersymmetric 
sine-Gordon theory.  

Work along these lines is in progress.

\section*{ACKNOWLEDGEMENTS}
We are grateful to Y. Billig, V. Frolov, N. Kaloper and R.
Myers for useful conversations.


\begin{thebibliography}{99}

\bibitem{hor}  For a review, see G. T. Horowitz,  `Quantum States of
Black Holes', UCSBTH preprint UCSBTH-97-06, gr-qc/9704072 (1997).


\bibitem{sol}  For a review, see {\it Solitons}, ed. R.K. Bullough and
P.J. Caudrey, Springer-Verlag, Berlin (1980).  For references to the
early work on sine-Gordon by Enneper, B$\ddot a$cklund, etc., see the
contribution to this volume by R.K. Bulloough.

\bibitem{cramp}  M. Crampin, et. al., Rep. on Math. Phys., {\bf 17}, 373
(1980).


\bibitem{jt}  See the contributions of R. Jackiw and C. Teitelboim in
{\it Quantum Theory of Gravity}, ed. by S.M. Christensen, Adam Hilger,
Bristol, (1984).


\bibitem{btz}  M. Banados,  C. Teitelboim and J. Zanelli,
Phys. Rev. Lett. {\bf69}, 1849 (1992); M. Banados,  M. Henneaux,
C. Teitelboim and  J. Zanelli, Phys. Rev. {\bf D48}, 1506 (1993).
M. Banados, `Constant Curvature Black Holes', preprint gr-qc/9703040 (1997).


\bibitem{bf}  T. Fukuyama and K. Kamimura, Phys. Lett. B{\bf 160}, 259
(1985);  K. Isler and C. Trugenberger, Phys. Rev. Lett. {\bf 63}, 834
(1989);  A. Chamseddine and D. Wyler, Phys. Lett. B{\bf 228}, 75
(1989); cf.\ also P. Schaller and T. Strobl, Phys. Lett. B{\bf 337},
266 (1994).

\bibitem{obs} J. Gegenberg, G. Kunstatter and D. Louis-Martinez, Phys.
Rev. {\bf D51}, 1781 (1995).

\bibitem{robb1}  R.B.Mann, Phys. Rev. {\bf D47},4438 (1993);  J.P.S
Lemos and P.M. Sa, Dept. de Fisica, Instituto Superior Tecnico
preprint DF/IST-9.93 (1993).

\bibitem{akns} M.J. Ablowitz, et. al. Stud. Appl. Math. {\bf 53}, 249
(1974).

\bibitem{witten}  E. Witten and D. Olive, Phys. Lett. B{\bf 78},97 (1978).

\bibitem{rebhan} A. Rebhan and P. van Nieuwenhuizen, `No saturation of the 
quantum Bogomolnyi bound by two-dimensional supersymmetric solitons', hep-th/
9707163.

\end{thebibliography}
\end{document}